\documentclass[lettersize,journal]{IEEEtran}
\usepackage{amsmath,amsfonts}
\usepackage{algorithmic}
\usepackage{array}
\usepackage[caption=false,font=normalsize,labelfont=sf,textfont=sf]{subfig}
\usepackage{textcomp}
\usepackage{stfloats}
\usepackage{url}
\usepackage{verbatim}
\usepackage{graphicx}

\usepackage[active]{srcltx}
\usepackage{color}

\def\BibTeX{{\rm B\kern-.05em{\sc i\kern-.025em b}\kern-.08em
    T\kern-.1667em\lower.7ex\hbox{E}\kern-.125emX}}
\usepackage{balance}

\begin{document}
\title{Reliability and Reproducibility of the Cryogenic Sapphire Oscillator Technology.}

\author{Christophe Fluhr$^\nabla$, Beno\^{i}t Dubois$^\nabla$, Guillaume Le Tetu$^\sharp$, Valerie Soumann$^\sharp$, \\ 
Julien Paris$^\flat$, Enrico Rubiola$^{\sharp\, \otimes}$ and Vincent Giordano$^\sharp$

\thanks{Manuscript created October 2022}
\thanks{$^\nabla$ France Comt\'{e} Innov, 25000 Besan\c{c}on, France.}
\thanks{$^\sharp$ FEMTO-ST Institute, Dept.\ of Time and Frequency, Universit\'{e} de Bourgogne and Franche-Compt\'{e} (UBFC), and Centre National de la Recherche Scientifique(CNRS), E-mail: giordano@femto-st.fr,~~  
Address: ENSMM, 26 Rue de l'Epitaphe, 25000 Besan\c{c}on, France.}
\thanks{$^\flat$ My Cryo Firm, 94120 Fontenay-sous-Bois, France.}
\thanks{$^\otimes$ Physics Metrology Division, Istituto Nazionale di Ricerca Metrologica INRiM, Torino, Italy.}
}

\maketitle

\begin{abstract}
\boldmath
The cryogenic sapphire oscillator (CSO) is a highly specialized machine, which delivers a reference signal exhibiting the lowest frequency fluctuations. For the best units, the Allan deviation (ADEV) is $\sigma_{y}(\tau) < 10^{-15}$ for integration time between $1$ and $10^{4}$ s, with a drift $<10^{-14}$ in one day. The oscillator is based on a sapphire monocrystal resonating at 10 GHz in a whispering-gallery mode, cooled at $6~$K for highest Q-factor and zero thermal coefficient. We report on the progress accomplished implementing eleven CSOs in about 10 years since the first sample delivered to the ESA station in Argentina.  Short-term stability is improved by a factor of 3-10, depending on $\tau$, and the refrigerator’s electric power is reduced to 3 kW.  Frequency stability and overall performances are reproducible, with unattended operation between scheduled maintenance every two years.
The CSO is suitable to scientific applications requiring extreme frequency stability with reliable long-term operation.  For example, the flywheel for primary frequency standards, the ground segment of GNSS, astrometry, VLBI, and radio astronomy stations.
\unboldmath
\end{abstract}


\section{Introduction}

\IEEEPARstart{S}{hort} term fractional frequency stability in the $10^{-15}$ range has been demonstrated more than 20 years ago with the use of high Q-factor microwave dielectric sapphire resonator cooled near the liquid helium temperature \cite{luiten94,wang99}. In the early 2000s, Cryogenic Sapphire Oscillator (CSO) breakthrough performances and early uses in the field of Time and Frequency Metrology have been demonstrated with prototypes still operating within a liquid He bath \cite{santarelli99-prl,chang00,wolf03,ell04-lowdrift,uffc04_open_cavity}. In 2010, at the FEMTO-ST Institute, we demonstrated for the first time the possibility to use a cryocooler while maintaining a state-of-the-art frequency stability \cite{rsi10-elisa}. Since, we undertaken large engineering efforts to rationalize the CSO design and its development, reduce its electrical consumption and improve its immunity to environmental perturbations \cite{uffc-2011-long-term-stab,jap-2014,cryogenics-2016,jap-2020}. \\

Currently, there is no competing technology.  The closest are the laser stabilized to a cryogenic Fabry-Perot cavity and the Hydrogen maser, but they do not show the best stability in same region of the $\sigma_{y}(\tau)$ plot, and have quite different ``personalities''.	 
The short term frequency stability achieved by the CSO is today only surpassed by some high-class laser prototypes stabilized on a cryogenic ultra-stable Fabry-Perot cavity \cite{kessler2012,robinson2019}. Commercial versions based on a room temperature Ultra-Low Expansion (ULE) optical cavity approach the short-term CSO frequency stability but still present a large long term frequency drift, i.e. $\sim 10^{-11}-10^{-12}$/day \cite{www.menlo-ors,www.stable-laser-systems}. Moreover, those optical sources operate generally in the near IR, and then require a metrological femto-second laser to derive useful signals in the microwave or the VHF bands. \\

Our upgraded CSO technology, code named ULISS-2G, is today sufficiently mature to be offered as a commercial product consuming only 3 kW and able to run continuously for year. A simple maintenance operation, which can be performed by the user, is only required every two years. \\

In this paper, we present the performance as well as the operating history of all the CSOs that we have built and validated since 2009. The objective here is to demonstrate the reliability and the reproducibility of our CSO technology. \\
 
\section{FEMTO-ST CSO design}
\begin{figure}[h]
\centering
\includegraphics[width=0.85\columnwidth]{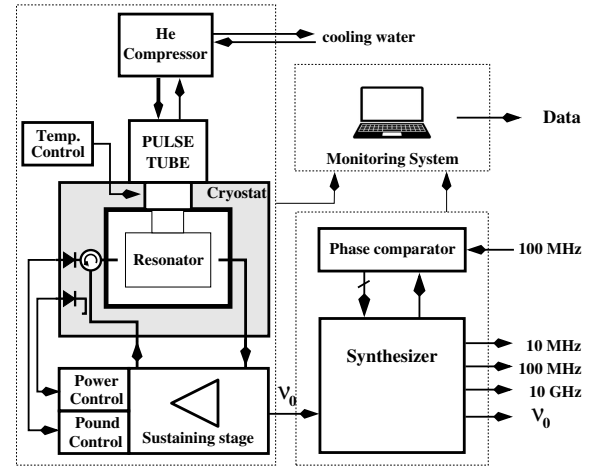}
\caption{CSO block diagram: we distinguish 3 mains subsets: the ultra-stable oscillator itself, the frequency synthesizer and the monitoring system.}
\label{fig:fig1}
\end{figure}

\noindent A block diagram of the CSO is represented in the figure \ref{fig:fig1}. 

\noindent The high Q sapphire resonator is maintained near $6$~K in a cryostat cooled with an autonomous Pulse-Tube (PT) cryocooler. The CSO is a Pound-Galani oscillator: the resonator is used in transmission mode in a regular oscillator loop, and in reflection mode as the discriminator of the classical Pound servo. The sustaining stage and the control electronics are placed at room temperature. The CSO output signal at the resonator frequency $\nu_{0}$ drives the frequency synthesizer, which eventually delivers several output frequencies: $10$~GHz, $100$~MHz and $10$~MHz in the typical implementation. Eventually the synthesizer outputs can be disciplined at long term on an external 100 MHz signal coming from an Hydrogen Maser for example. A monitoring system scrutinizes all the key parameters to follow the CSO status.
The technological choices relating to the various subsets have already been described in previous publications to which we refer the reader  \cite{ell10-elisa,ell11-elisa,journal-physics-2016,uffc-2016}. Here we will simply recall the key features, which make the originality of our design.\\

\subsection{Sapphire resonator}

\begin{figure}[h]
\centering
\includegraphics[width=0.8\columnwidth]{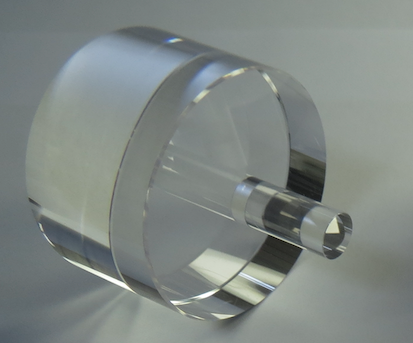}
\caption{Cryogenic sapphire resonator operating on the WGH$_{15,0,0}$ mode at 9.99 GHz.}
\label{fig:fig2}
\end{figure}

\noindent The cylindrical resonator 54-mm-diameter and 30-mm-height is made in a high purity sapphire monocrystal (see Fig. \ref{fig:fig2}). It operates on the quasi-transverse magnetic whispering-gallery mode WGH$_{15,0,0}$ at $\nu_{0}=9.99$~GHz~$\pm 5$~MHz. This choice greatly simplifies the design of the frequency synthesis (see below). The sapphire resonator frequency shows near 6 K a turnover temperature $T_{0}$ for which the resonator sensitivity to temperature variations nulls at first order. The appearance of this turning point results from the presence in Al$_{2}$O$_{3}$ of paramagnetic impurities as Cr$^{3+}$ or Mo$^{3+}$, whose concentration are typically well below $1$~ppm in weight. The exacte value of $T_{0}$ is specific to each resonator and for high quality sapphire crystal is typically found between  $5$~and $9$~K \cite{mtt-2015}. At $T_{0}$, the unloaded Q factor can achieve two billion depending on the sapphire crystal quality and on the resonator adjustment and cleaning.\\

The sapphire resonator includes an integral spindle (see Fig.~\ref{fig:fig2}), such that the sapphire piece can be mounted with no significant fixing stress in the cylinder's critical circumferential region, where the WG mode’s field is concentrated.
Since 2009, we have ordered more than 25 sapphire resonators from several high-quality crystal manufacturers that we selected after preliminary tests on samples. Of these resonators, $17$ were actually tested at low temperature. The following figures show the reproducibility of resonator main characteristics.\\

\subsubsection{$\nu_{0}$, resonator frequency} 
~
\begin{figure}[h]
\centering
\includegraphics[width=\columnwidth]{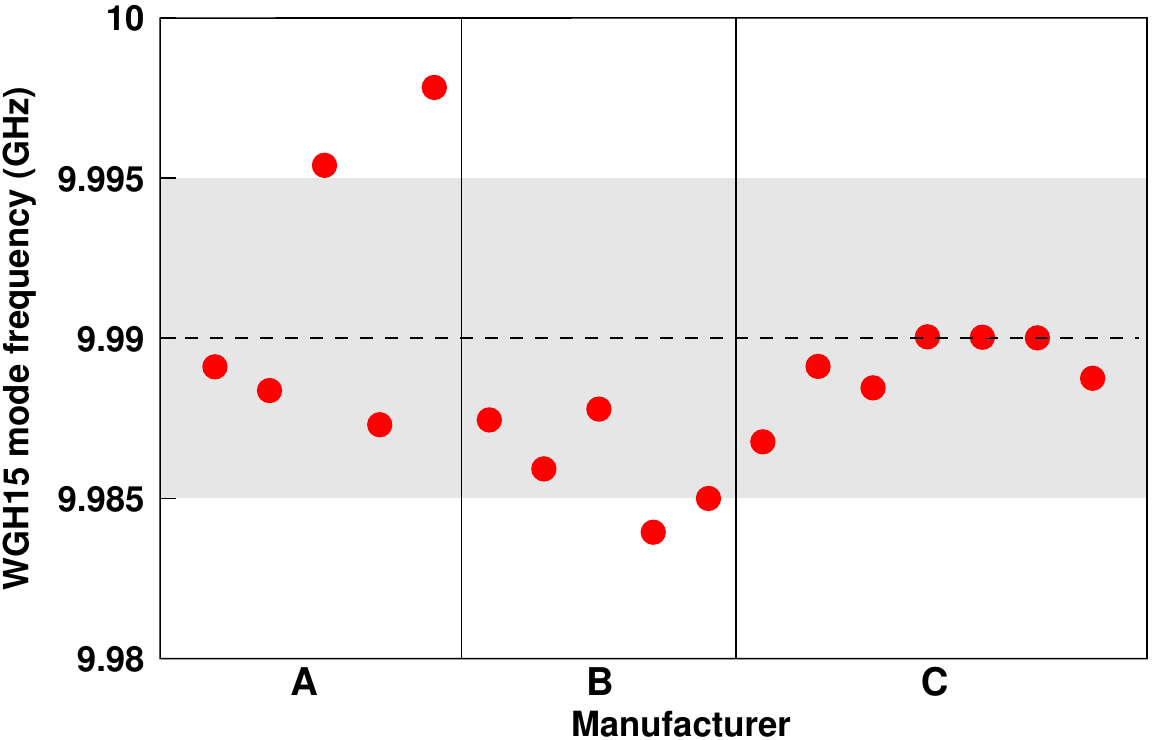}
\caption{WGH$_{15,0,0}$ mode frequency of the tested resonators.}
\label{fig:fig3}
\end{figure}

We had set the tolerances on resonator dimensions to achieve a frequency of $9.99$~GHz~$\pm
5$~MHz which simplifies the design of frequency synthesis (see Sect. \ref{sec:Frequency synthesis} ). Only one among the tested resonators presents a frequency notably shifted from the expected value at about 9.998 GHz. Nevertheless, it could be integrated into a CSO by making some minor modifications to the synthesis.\\

\subsubsection{$T_{0}$, resonator turnover temperature} 
~
\begin{figure}[h]
\centering
\includegraphics[width=\columnwidth]{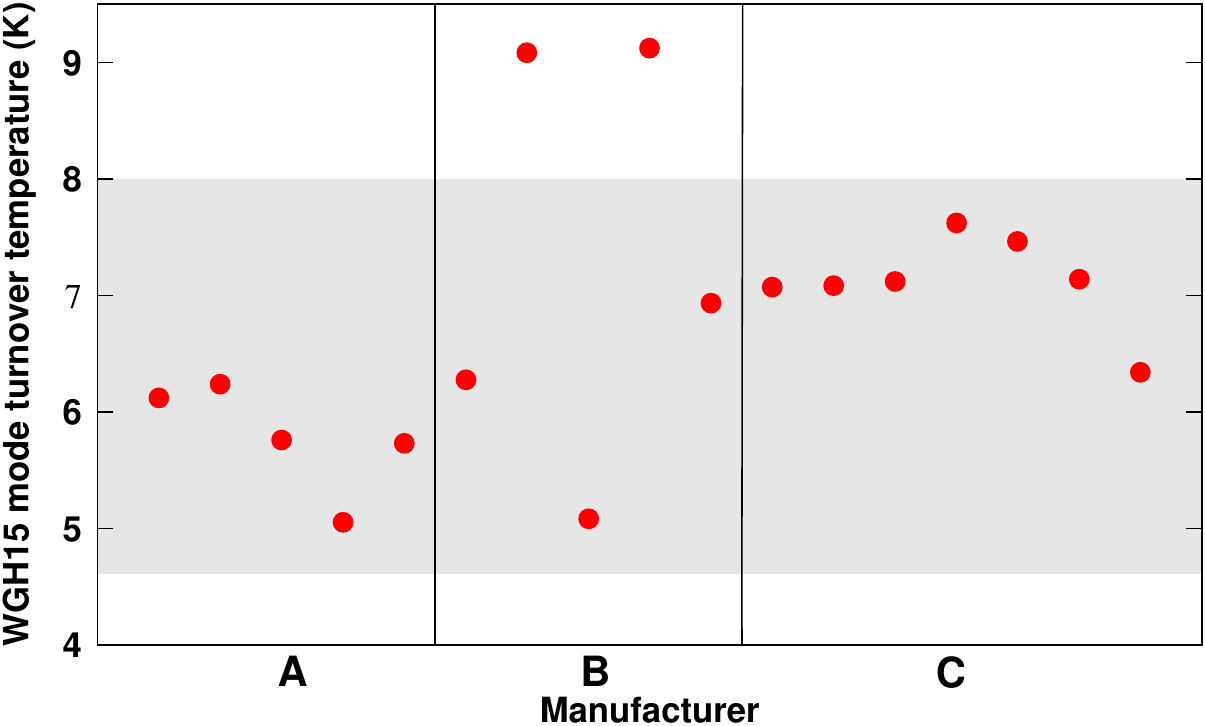}
\caption{WGH$_{15,0,0}$ mode turnover temperature of the tested resonators. The grey zone: $4.6~$K$\leq T_{0}\leq 8~$K corresponds to our specification.}
\label{fig:fig4}
\end{figure}

The lowest value of the acceptable temperature range is $4.6$~K, which is the lowest achievable temperature in ULISS-2G increased by 0.5 K. It represents the lowest allowable $T_{0}$ for an effective temperature stabilization.
On the other hand, when $T_{0} \geq 8~$K, we will see in the section \ref{sec:uliss-2G} that the CSO's ADEV is degraded at short term and becomes higher than $3\times 10^{-15}$.\\

\subsubsection{$Q_{0}$, resonator unloaded Q-factor} 

We set the lowest allowable unloaded Q-factor to $5\times 10^{8}$. With $Q_{0}>5\times 10^{8}$, the overall effect on the frequency stability of the noise associated with the oscillator electronics is kept below $1\times 10^{-15}$ \cite{mtt-2015}. All the tested resonators fulfil this requirement. The achievement of a high unloaded Q-factor is not only dependent on the crystal quality, but also on the resonator cleaning. We developed our own cleaning method realized in a class 100 clean-room in order to suppress any particule that could stick on the resonator surface.\\

\begin{figure} 
\centering
\includegraphics[width=\columnwidth]{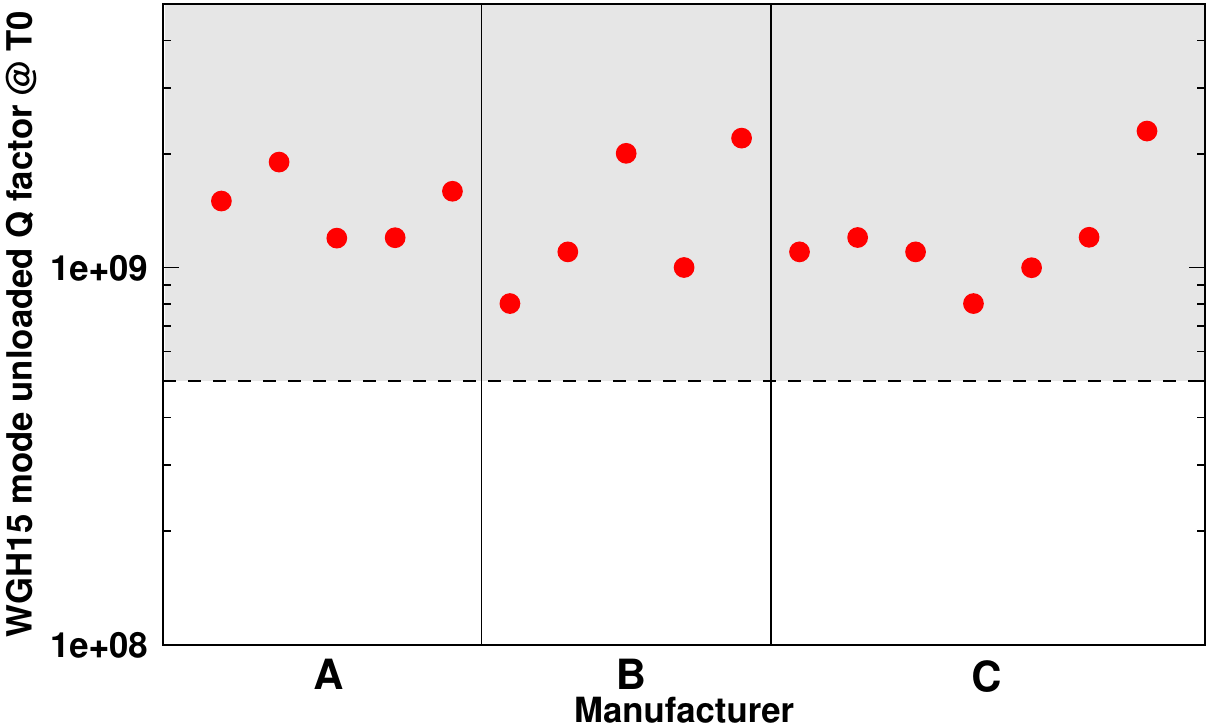}
\caption{WGH$_{15,0,0}$ mode unloaded Q-factor of the tested resonators.}
\label{fig:fig5}
\end{figure}

\begin{figure}[h]
\centering
\includegraphics[width=\columnwidth]{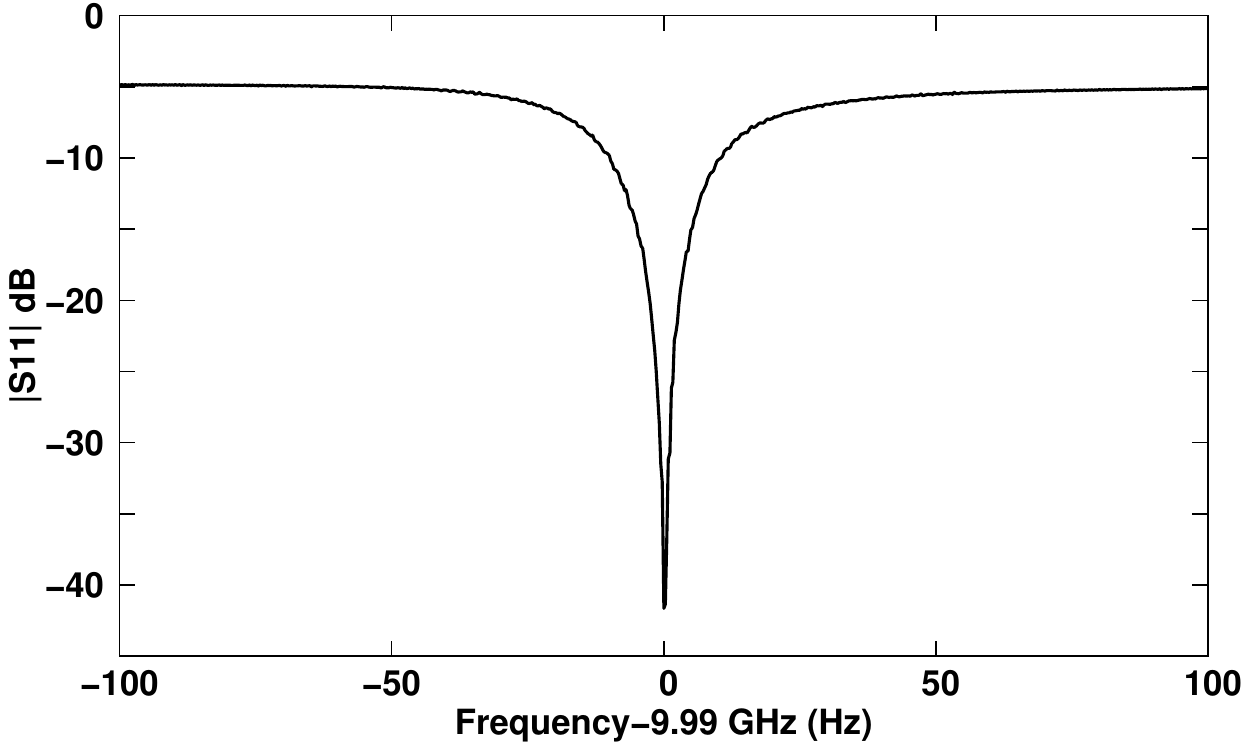}
\caption{Cryogenic sapphire resonator reflexion coefficient at 9.99 GHz}
\label{fig:fig6}
\end{figure}

The resonator is placed in the center of an OFHC copper cylindrical cavity and the resonator coupling is realized through two magnetic loops protubating inside the cavity. To optimize the oscillator operation, the cooled resonator should be critically coupled at its input port, meaning that at resonance the resonator input reflexion coefficient should be $S_{11}(\nu_{0})\approx 0$. The difficulty is that the resonator coupling behave as its Q-factor, and thus is multiplied by some 10,000 when the temperature is decreased down to $6~$K. We developped a specific procedure to adjust at room temperature the resonator input reflexion coefficient to get the optimum coupling at $T_{0}$. This procedure requires only two successive cool-downs. The figure \ref{fig:fig6} shows the magnitude of the cryogenic resonator input port reflexion coefficient  obtained after the coupling adjustment showing a nearly perfect critically coupled input.\\

\subsection{Sustaining loop}
\label{sec:Sustaining loop}
\noindent 
The amplifying stage, which operates at room temperature, has a 100 MHz input bandwidth fixed by an internal bandpass filter centered on $9.99~$GHz. Its gain and phase lag are set independently by two external near DC voltages, which constitue the actuator inputs for the power and Pound frequency controls. The 100 kHz phase modulation, needed for the Pound frequency lock, is applied to the output signal that is injected in the resonator.  
An internal coupler derives a part of the loop signal to get the $10 \pm 1$~dBm reference signal driving the frequency synthesis.  \\

More critical are the components placed near the resonator at low temperature. Microwave isolators and circulator are commercial SMA connectorized components that we selectionned after low temperature tests and cycling. The more sensible components are the two tunnel diodes that are used for the Power and Pound controls \cite{rsi-2016}. In order to avoid any degradation of these sensitive detectors, we intentionnaly limit the power injected in the cryostat to $-5$~dBm. \\

In some CSO designs \cite{locke08}, the frequency stability is affected by amplitude modulation (AM)-index fluctuations of the Pound interrogation signal. Such a sensitivity makes mandatory a supplementary control loop to suppress the spurious AM. In our CSOs, we do not implement such a AM suppression lock loop as the nearly perfect critical coupling of the resonator makes this phenomena non relevant.

\subsection{Frequency synthesis}
\label{sec:Frequency synthesis}
\begin{figure}[h]
\centering
\includegraphics[width=0.99\columnwidth]{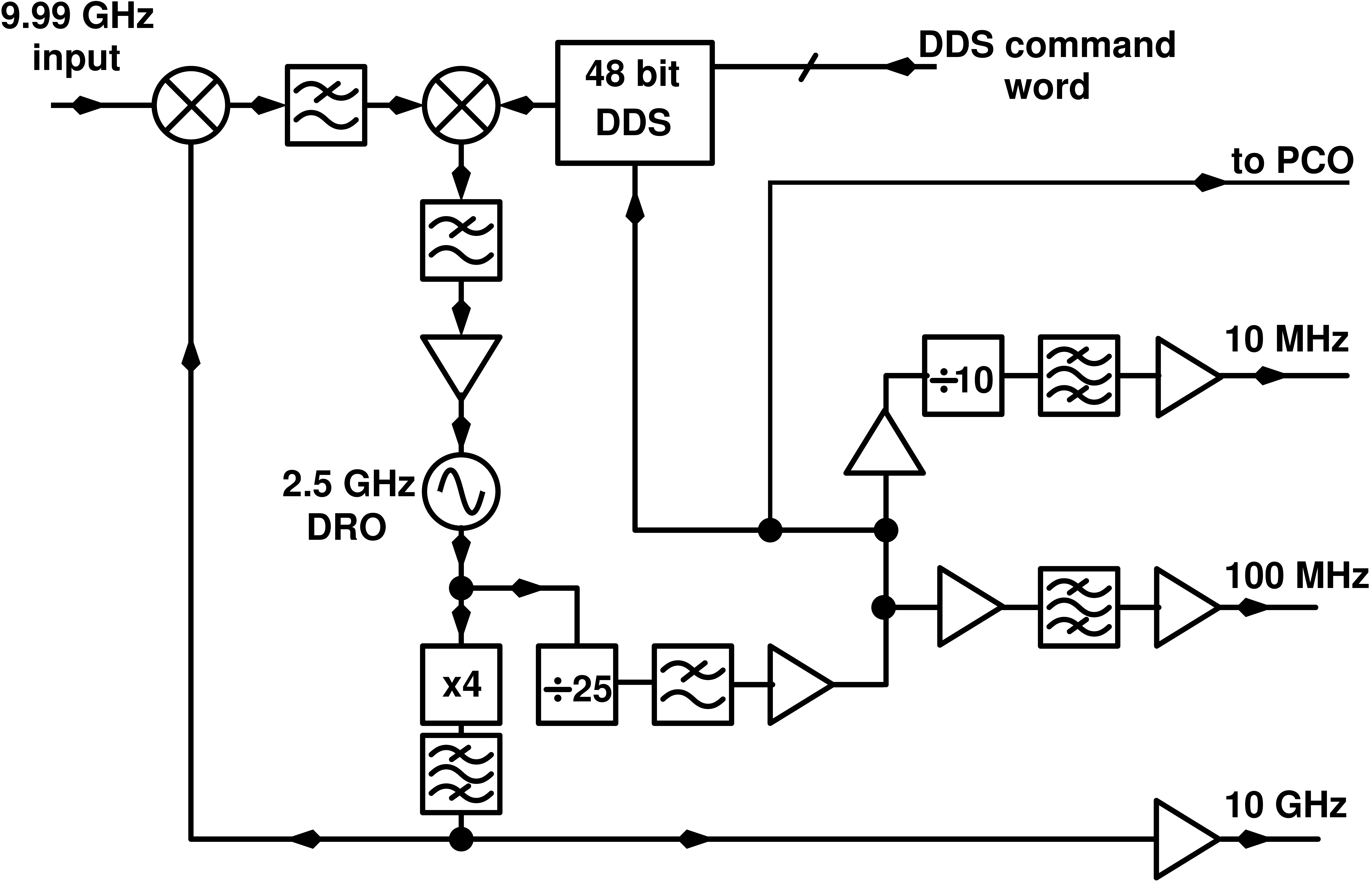}
\caption{ULISS-2G frequency synthesizer. VCO: voltage controled frequency multiplied Dielectric Resonator Oscillator (DRO). DDS: Direct Digital Synthesizer. PCO: $100$ MHz Phase COmparator.}
\label{fig:fig7}
\end{figure}

\noindent 
The frequency synthesizer principle is shown in the figure \ref{fig:fig7}.

A $2.5$~GHz DRO selected for its low phase noise is frequency multiplied by $4$ and then, phase locked on the CSO output signal. The frequency difference, i.e. $10~$MHz~$\pm 5~$MHz , is compensated by a low noise $48$~bit DDS enabling a $10^{-16}$ fractional frequency resolution. The output $100~$MHz and $10$~MHz signals are generated by simple frequency divisions. The generated $100~$MHz can also be send to the Phase Comparator (PCO) driven by  an external reference signal. A digital control word is thus obtained to control the DDS frequency and disciplinate at long term the CSO synthesizer output signal.

\subsection{CSO production history}
\noindent 
Since 2009, we built and validated eleven CSOs. Our first cryocooled CSO was build in the frame of the ELISA project, funded by the European Space Agency. In 2010, ELISA demonstrated for the first time a state-of-the-art frequency stability with an autonomous cryocooler instead of a liquid helium bath \cite{rsi10-elisa}.  As we will see in the next sections, our technology evolved in 2015 with the introduction of a low-power cryocooler. The table \ref{tab:tab-cso} presents these CSOs, giving the electrical consumption and  the date of the first operation.
 
\begin{table*}
\centering
\caption{The 11 CSOs built and validated at the FEMTO-ST Institute.}
\label{tab:tab-cso}
\begin{tabular}{ccccl}
\hline
\#	& Nickname	& Power	& first operation & status \\
\hline
000	& ELISA			& 6 kW	& June 2009	& Prototype \\
001	& MARMOTTE		& 6 kW	& Oct. 2010	& OSCILLATOR IMP Reference \\
002	& ULISS			&6 kW	& Nov.  2011	& OSCILLATOR IMP Reference, Transportable unit\\
003	&  ABSOLUT		& 7 kW	& May  2014	& OSCILLATOR IMP Reference\\
004	& ULISS-2G proto	& 3 kW	& June 2015	& Prototype, principle demonstration \\
005	&ULISS-2G 005	& 3 kW	& Dec. 2017	& Commercial product (delivered)\\
006	&ULISS-2G 006	& 3 kW	& Nov. 2018	& Commercial product (delivered)\\
007	&ULISS-2G 007	& 3 kW	& June 2019	& Commercial product (delivered)\\
008	&ULISS-2G 008	& 3 kW	& Dec. 2020	& Commercial product (delivered)\\
009	&ULISS-2G 009	& 3 kW	& April 2021	& Commercial product (delivered)\\
010	&ULISS-2G 010	& 3 kW	& March 2022	& Commercial product (still in validation phase)\\
\hline
\end{tabular}
\end{table*}

\section{ULISS-1G: $1^{st}$~generation of crycooled CSO}
\label{sec:uliss-1G}
\noindent 

\begin{figure}[h]
\centering
\includegraphics[width=\columnwidth]{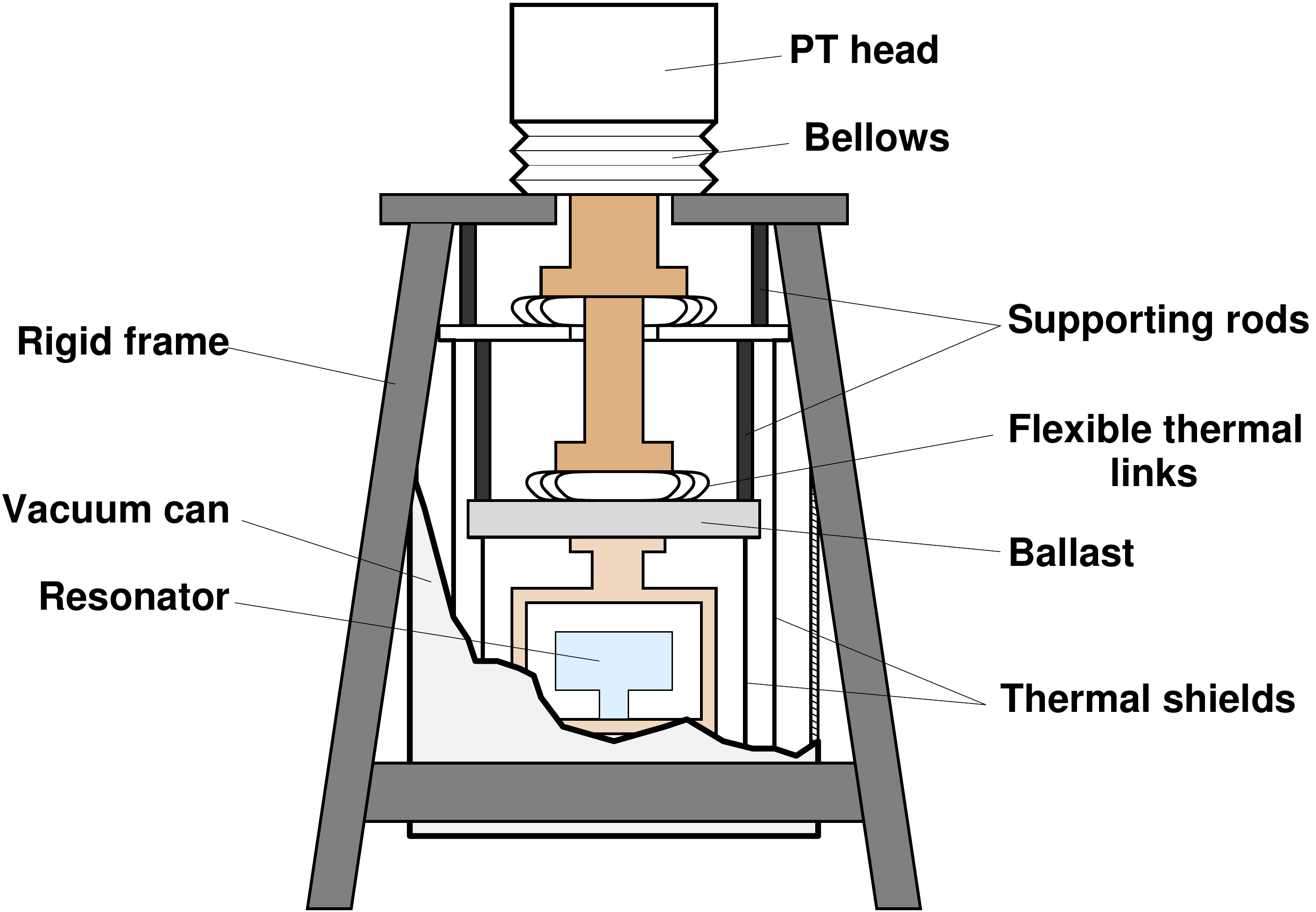}
\caption{ULISS-1G cryostat}
\label{fig:fig8}
\end{figure}

The tradeoff in the cryostat design is to ensure a proper thermal conduction between the resonator and the cold source, while limiting the transfer of mechanical vibrations arising from the cryocooler. The ULISS-1G cryostat principle is given in the figure \ref{fig:fig8}.  We opted for simple solutions, favoring passive thermal filtering and mechanical decoupling by flexible links. The resonator is supported by stiff rods attached to a rigid frame whereas it is linked to the cold source by a set of copper braids. The fundamental frequency of the PT thermal cycle is typically $\sim 1.4$~Hz. At such a low frequency, the transfer of the cold finger mechanical displacement to the resonator is attenuted by the ratio of the copper braids stiffness to those of the supporting rods.  Moreover, the temperature variations observed on the cold finger are passively filtered by the thermal ballast constituted by the stainless steel top flange of the 2nd stage thermal shield. Associated with the thermal resistance of the flexible thermal link it is equivalent to a first order filter.

For ULISS-1G we selected powerful cryocoolers providing at least $500$~mW of cooling power at $4~$K. With such a cooling power, the constraints in the material and geometry of the supporting rods and in the residual stiffness of the flexible thermal links are greatly relaxed. Thus, a minimal temperature of $4~$K and a residual displacement below $1~\mu$m at the resonator level are easily reached.

ELISA was validated in 2010 and later moved to the ESA Deep Space Antenna Station DSA-3 in Malarg\"ue, Argentina. Thus all the instrument was conveyed by air to Buenos Aeres, then by truck to Malargue with about 30 km of an unpaved track to get eventually the ESA station. Vibrations and schocks occuring during this long transportation did not affect the resonator adjustment, and ELISA was put into operation in only two days after its arrival.\\

\begin{figure}[!h]
\centering
\includegraphics[width=\columnwidth]{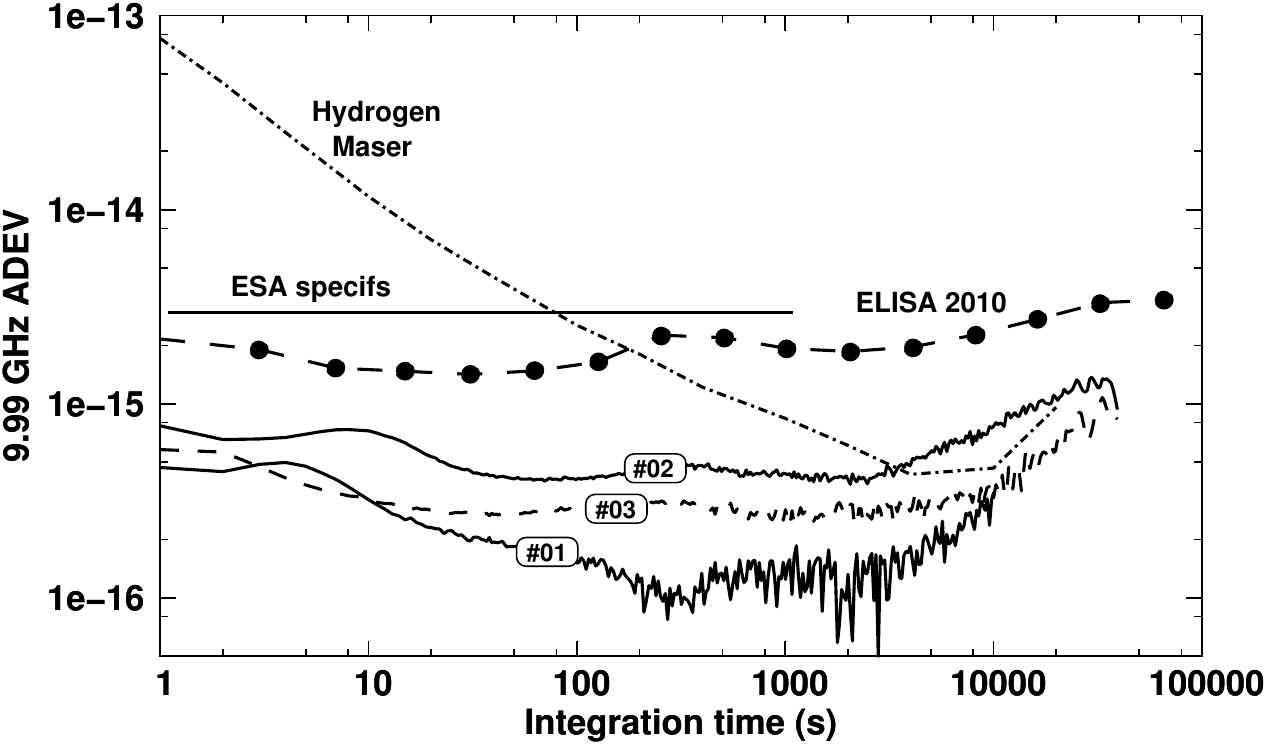}
\caption{Fractional frequency stability of the ULISS-1G CSOs. ELISA was the first prototype built for ESA and now in Malargue (Ar). The other 3 units are integrated in the OSCILLATOR-IMP platform, thus providing the microwave reference for short-term frequency stability. For comparison the ADEV of a high performances Hydrogen Maser is also represented.}
\label{fig:fig9}
\end{figure}

After the success of the ELISA project, we decided to build three other units to be integrated in our metrological platform OSCILLATOR-IMP \cite{www.osc-imp}. These 3 CSOs were implemented successively between 2010 and 2014 in a temperature stabilized room at $22~^{\circ}$C $\pm0.5~^{\circ}$C. Since then, they have been running almost continuously. Both Pound and power servo loops are very stable and robust. No accidental loss of control has ever been recorded. Several short stoppings were carried out to enable i) their routine maintenance: one He filter should be changed every 2 years in the compressor circuit, ii) the maintenance of the building electric circuit. The CSOs are also stopped for energy saving reasons during the two weeks of summer holidays when the laboratory is closed. We also recorded two breakdowns: one CSO Rotary Valve and one cryogenic power detector. These defective elements were quickly replaced by new ones, and the faulty CSOs recovered rapidly their initial performances. The figure \ref{fig:fig9} shows the fractionnal frequency stability of the ULISS-1G CSOs. \\

The best unit presents a short term frequency stability of $5\times10^{-16}\tau^{-1/2}$ limited by a flicker floor of $~1.5\times 10^{-16}$ between $100~$s and $5,000~$s. \\

\begin{figure}[!h]
\centering
\includegraphics[width=0.8\columnwidth]{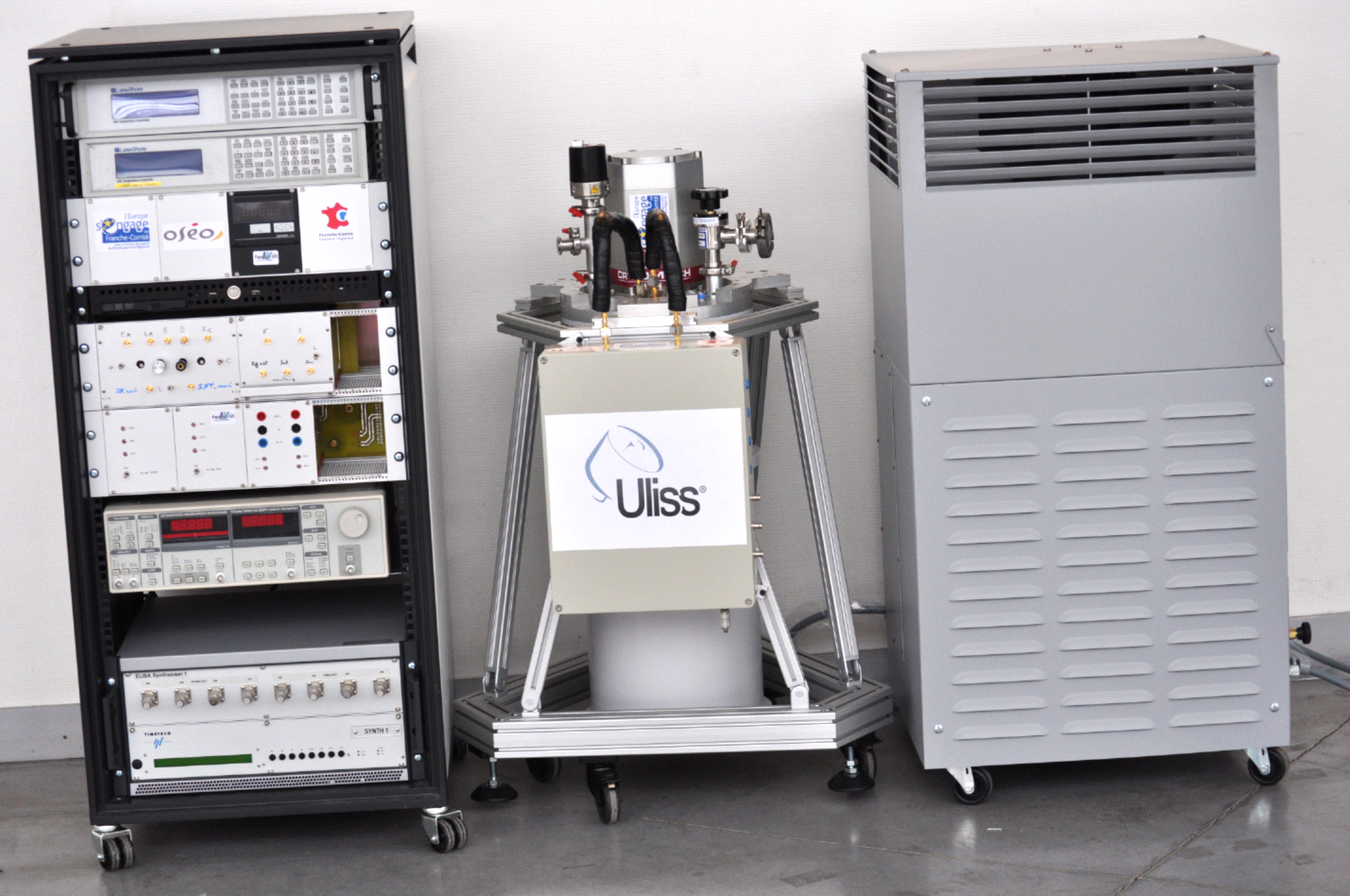}
\caption{The CSO of $1^{\mathrm{st}}$ generation: ULISS-1G. From the left: 19'' rack including the electronic and temperature servos as well as the frequency synthesis, the cryostat with the sustaining stage in the front box, and the air cooled He compressor. This unit was designed to be transported in a small van to be tested in different sites around Europe. In Femto-ST, the CSO are running with a water cooled compressor.}
\label{fig:fig10}
\end{figure}

On of these CSOs, see the figure \ref{fig:fig10}, was specifically designed to be transported in a small truck, with the aim of testing the CSO technology in some European metrological sites to evaluate its potential for real applications. In two years we visited several European sites, traveling more than 10,000 km \cite{rsi-2012}, without any problem related to the transportation, demonstrating the robustness of our design.

The ULISS \it{Odissey}\rm\ was the opportunity to meet potential users and understand their needs and possible difficulties related to the implementation of the CSO in their laboratory \cite{grop2012uliss,AppPhysB-2014,abgrall2016}. It was pointed out that the electrical consumption (three-phase 8 kW peak/6 kW stable operation) and the heat released by the compressor could turn out to hinder the deployment of the CSO technology.

\section{ULISS-2G: $2^{nd}$ generation of crycooled CSO.}
\label{sec:uliss-2G}
\noindent 

\begin{figure}[!h]
\centering
\includegraphics[width=\columnwidth]{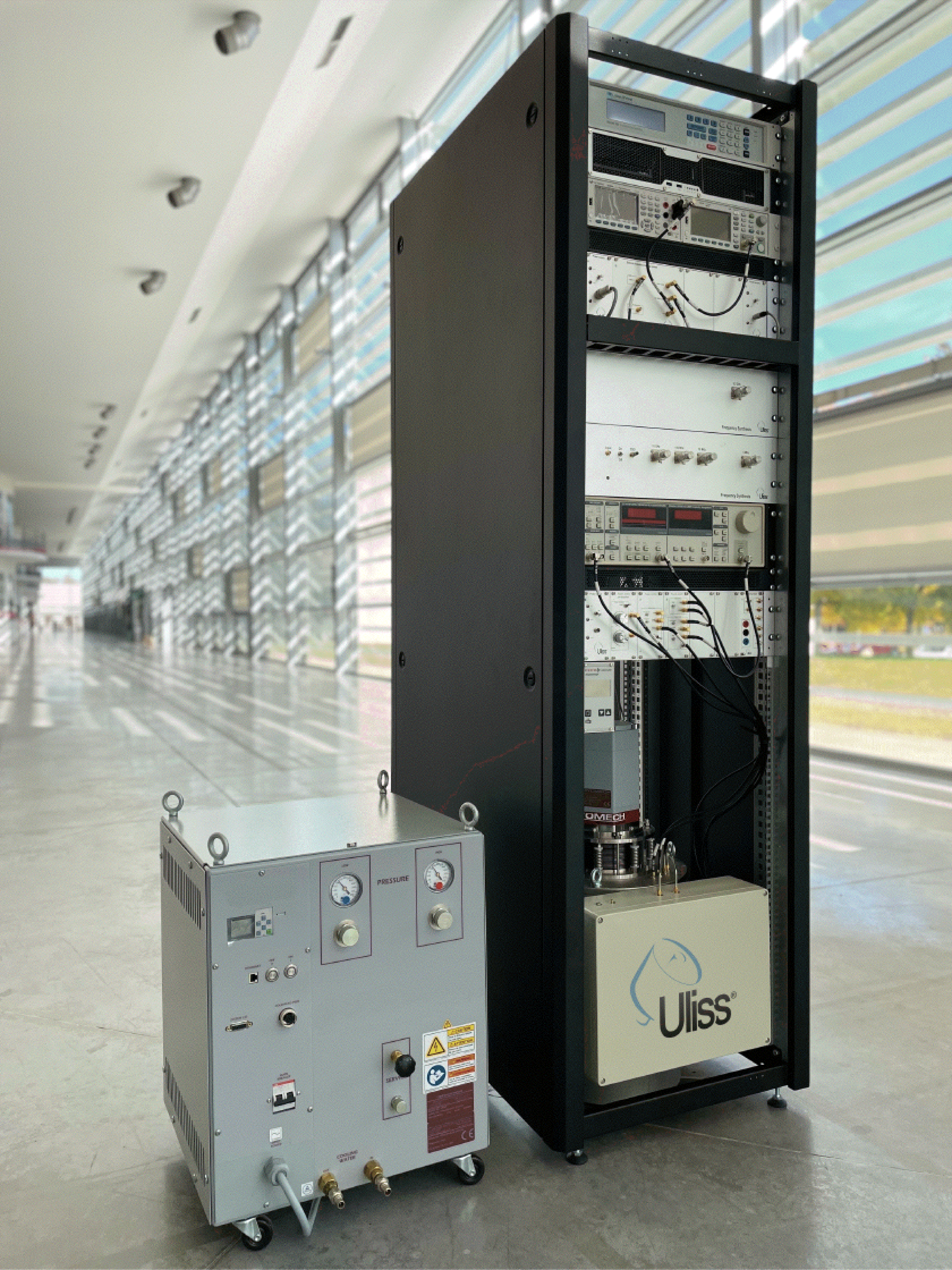}
\caption{The CSO of 2nd generation: ULISS-2G. Its electrical consumption is $3~$kW single phase.}
\label{fig:fig11}
\end{figure}

The ULISS-2G project was launched to solve the above mentionned technological difficulty with the aim to divide by two the electrical consumption. The cryostat was thus optimized to be able to maintain a minimum temperature of 
$\sim4~$K, with a cooling power of only $250$~mW, corresponding to the specification of a $3~$kW single phase Pulse-Tube PT403 from Cryomech \cite{www.cryomech}. This was achieved by improving the thermal shieldings, the conductance and flexibility of the copper braids and by designing new supporting rods realized in Mylar by 3D-printing. The new design of the cryostat was also the opportunity to rationalize its overall size: the supporting frame was suppressed and replaced by the vacuum can itself, which has been rigidified \cite{cryogenics-2016}. This allowed us to integrate the cryostat into a 19'' rack that also supports all the control electronics as well as the synthesis chain. The figure \ref{fig:fig11} shows the ULISS-2G instrument with its water cooled compressor.

\begin{figure*}[h]
\centering
\includegraphics[width=\textwidth]{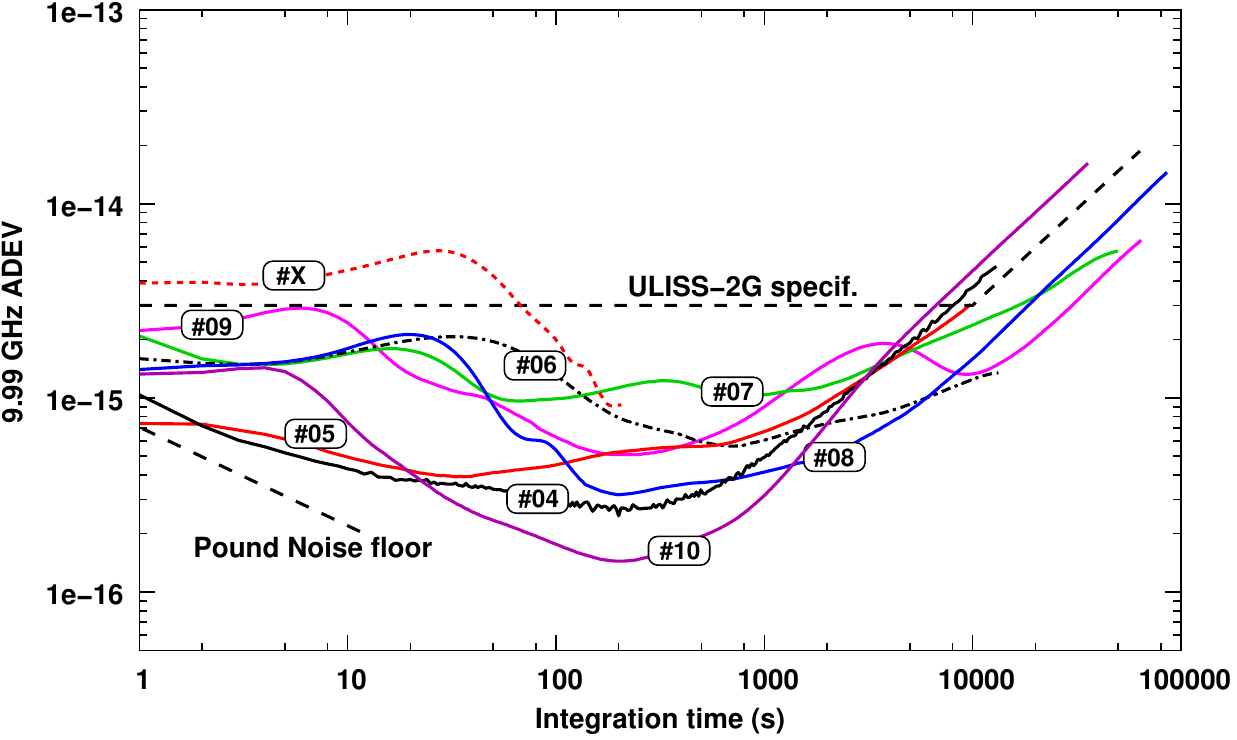}
\caption{Fractional frequency stability of the ULISS-2G CSOs: \#04 is the first prototype described in \cite{cryogenics-2016}. \#X has been obtained with a non-compliant resonator with $T_{0}\sim 9$~K. This resonator was eventually not retained.}
\label{fig:fig12}
\end{figure*}

Since 2017, we build and validated six ULISS-2G CSOs of identical design. The only notable difference between them is the value of the resonator turnover temperature $T_{0}$. The figure \ref{fig:fig12} shows the Allan deviation of these CSOs. These measurement results have been obtained by the three-cornered-hat-method with two of the OSCILLATOR-IMP platform CSOs as references. For all the ULISS-2G CSOs, the Allan deviation is better than $3\times10^{-15}$ for $1~\mathrm{s} \leq \tau \leq 10,000~\mathrm{s}$ and bellow $1\times10^{-14}$ over 1 day.\\

At short term, i.e. for  $1~\mathrm{s} \leq \tau \leq 100~\mathrm{s}$, none of the CSOs is limited by the  intrinsic noise of the Pound frequency discriminator, which is always below $7\times 10^{-15} \tau^{-1/2}$. Here, the CSO frequency stability is limited by the residual temperature fluctuations affecting the resonator. In \cite{uffc-2016}, we shown how unexpected time lags in the thermal system and the relative slowness of the temperature controler digital processing induce a slight and slow resonator temperature modulation. This modulation affects the short term fractional frequency stability, making a hump in the ADEV curve. This hump is clearly visible in the figure \ref{fig:fig12} between $5$~s and $100$~s, for five of the CSOs. Our confidence in this diagnostic is still reenforced by the fact that the best two ULISS-2G CSOs, i.e. \#04 and \#05, are characterized by a resonator turnover temperature of $5.1$~K and $5.2$~K, whereas for the next four, we have: $6.9~$K$< T_{0}<7.3~$K. \#X was obtained with a non-compliant resonator characterized by $T_{0}=9.0$~K. Despite its high Q-factor $\sim 10^{9}$, the short term ADEV is higher than the ULISS-2G specification, i.e. $\sigma_{y}(\tau) \leq 3\times 10^{-15}$. This resonator was not retained for delivery. The last implemented CSO, i.e. \#10, has been runing for the first time in March 2022 and is still under test at FEMTO-ST. Its turnover temperature is $6.2$~K and it shows a very good mid-term frequency stability approching $1\times 10^{-16}$ around $200$~s. The differences in performance between these nearly identical CSOs can be explained by considering the temperature sensitivity of the following parameters:\\

-\it{Residual resonator thermal sensitivity}\rm. Near $T_{0}$, the resonator frequency variation is well approximated by a quadratic law: $ \nu(T) \approx \nu(T_{0})(1+a(T-T_{0})^{2})$, with $a<0$ the curvature  in K$^{-2}$ . Thus, the residual resonator thermal sensitivity is:
\begin{equation}
\dfrac{1}{\nu_{0}} \dfrac{\Delta \nu}{\Delta T}=2a(T-T_{0})
\end{equation}
$a$ depends on $T_{0}$ and in the considered temperature range: $a \propto T_{0}^{2}$. It is thus multiplied by a factor of two between $5~$K and $7~$K. 

-\it{Temperature sensor sensibility}\rm. For zirconium nitride thin-film resistor temperature sensors used in our CSOs, the sensitivity in $\Omega /$K is typically divided by about $5$ between $5~$K and $7~$K \cite{courts2003review}.

-\it{Material physical properties}\rm. The heat conductance $k$ (W.m$^{-1}$.K$^{-1}$) and the heat capacity $C_{p}$ (J.kg$^{-1}$.K$^{-1}$) vary substantially between $5~$K and $7~$K, for all materials \cite{sciver2012low}. For copper,  $k \propto T$, which is of less importance, but $C_{p}$ is doubled from  $5~$K to $7~$K. For sapphire, $k$ and $C_{p}$ are both proportionnal to $T^{3}$.

All these phenomena make the lowest turnover temperature more favorable: the resonator sensitivity is lower, the thermal time constants should be minimized and the temperature sensor more sensitive. The thermal control should thus be more efficient.\\

 A long term, i.e. for $\tau >100~$s, the CSO frequency stability is mainly limited by the ambient temperature variations. For slow temperature variations, we estimated the overall temperature sensitivity of the instrument at $1\times 10^{-14}$~K$^{-1}$. The  CSO under test was installed in a workshop equipped only with the standard air-conditioning system of the flat. Depending on the sunlight the temperature near the cryostat can vary of several degrees during the day. Moreover the workshop is in free access for the laboratory staff and this makes it impossible to maintain an undisturbed ambient for the duration of the measurement (few days).

\section{Summary}
\noindent 
In this paper, we reported the frequency stability performances of the dozen of cryogenic sapphire oscillators we have built since 2009. Our most advanced technology, i.e. ULISS-2G, provides a fractional frequency stability better than $3\times 10^{-15}$ for $1~$s$<\tau<10,000~$s and below $1\times 10^{-14}$ over 1 day, with a limited electric consumption, i.e. $3$~kW single phase. These performances have been obtained in a standard laboratory environment. Implemented in a metrological room with an efficient temperature control and a limited human presence, such ultra-stable oscillators are able to reach a fractional frequency stability approaching $1\times 10^{-16}$ near $\tau=100~$s. Some of the CSOs have been put into operation few years ago, and continue to run without any noticeable aging.

\section{Acknowledgments}
\noindent
The work has been realized in the frame of the ANR projects: EquipEX Oscillator-Imp and LabEX First-TF. The authors would like to thank the Council of the R/'egion de Franche-Comt\'e for its support to the Projets d’Investissements d'Avenir.

\bibliographystyle{ieeetr}


\end{document}